\documentclass[a4paper,10pt]{extarticle}

\usepackage[utf8x]{inputenc}
\usepackage{amsfonts}
\usepackage{amsmath}
\usepackage{color}
\usepackage{subfig}
\usepackage{amsthm}
\usepackage{float}
\usepackage{stfloats}
\usepackage{graphicx}
\usepackage{xcolor}
\usepackage{hyperref}
\usepackage{vmargin}
\usepackage[toc,page,header]{appendix}
\usepackage[noblocks]{authblk}
\hypersetup{colorlinks=true,citecolor=blue,linkcolor=magenta}
\usepackage[parfill]{parskip}
\usepackage{changebar}
\date{}
\title{\bf Dynamics of Stochastic Epidemics on Heterogeneous Networks}
\author[1,*]{Matthew Graham}
\author[1,2]{Thomas House}
\affil[1]{Complexity Science, University of Warwick, Gibbet Hill
Road, Coventry, CV4 7AL, UK.}
\affil[2]{Warwick Mathematics Institute, University of Warwick, Gibbet Hill
Road, Coventry, CV4 7AL, UK.}
\affil[*]{Corresponding Author: Matthew.Graham@warwick.ac.uk}

\begin{document}

\maketitle

\begin{abstract}
\noindent{}Epidemic models currently play a central role in our attempts to
understand and control infectious diseases. Here, we derive a model for the
diffusion limit of stochastic susceptible-infectious-removed (SIR) epidemic
dynamics on a heterogeneous network.  Using this, we consider analytically the
early asymptotic exponential growth phase of such epidemics, showing how the
higher order moments of the network degree distribution enter into the
stochastic behaviour of the epidemic. We find that the first three moments of
the network degree distribution are needed to specify the variance in disease
prevalence fully, meaning that the skewness of the degree distribution affects
the variance of the prevalence of infection. We compare these asymptotic
results to simulation and find a close agreement for city-sized populations.
\end{abstract}


\section{Introduction}
\setcounter{changebargrey}{0}
Epidemic models take a variety of mathematical
forms~\cite{Anderson&May1992,Keeling&Rohani2008}, and are are now routinely
used to inform policy on disease control and contribute towards public health
plans~\cite{Ferguson:2003,Riley:2003,Tildesley:2006,Baguelin:2010}.  One of the
most commonly investigated forms of models are susceptible-infectious-removed
(SIR), in which everyone is either susceptible to the disease, infectious with
it, or removed from the future disease dynamics. This model is a simplification
of reality, but provides a useful starting point for modelling diseases where
previous infection confers long-lasting immunity, for example outbreaks of
childhood diseases like measles, some respiratory illnesses like pandemic
influenza, and historical pathogens such as
smallpox~\cite{Keeling&Rohani2008}.

Simple models of epidemics have assumed that all members of a population
interact at a homogeneous rate and therefore that at a given time every
susceptible has an equal probability of contracting the disease from any
infectious individual.  A more realistic way to think of the spread of diseases
can be that they spread through contacts between people, with these contacts
describing a network of interactions. In the case of homogeneous mixing this
network is under some conditions an Erd\"{o}s-R\'{e}nyi (ER) random graph.
There have been many examples of using networks to study the spread of disease
and two review papers~\cite{Bansal2007,Danon:2011} compare several different
approaches to network modelling.

Longstanding generalisations of homogeneous mixing epidemic dynamics include
heterogeneous network models~\cite{Eames:2002} and regular clustered
networks~\cite{Keeling1999}.  Regularity (each individual having exactly $n$
links) is appropriate for some populations (and was originally motivated by
spatially embedded ecological and veterinary applications) but for human
respiratory transmission~\cite{Mossong:2008} and sexually transmitted
infections~\cite{Schneeberger:2004,Kamp:2010}, this is almost definitely
inaccurate and the contact network is highly degree-heterogeneous.  In this
work, we represent heterogeneity in degree using what has become a \textit{de
facto} standard: the configuration model~\cite{MolloyReed}.

As relevant theory has developed, through moment-closure approximations and
other methods~\cite{ARand1999,Bansal2007,Danon:2011,Lindquist2011}, the use of
heterogeneous networks has become more and more common in the literature.
Several areas of interest for epidemics have been studied. The invasion
threshold of the infection has been calculated~\cite{Diekmann2000}, and
involves the mean and variance of the degree distribution.  The final size of
an epidemic with a given degree distribution along with the mean degree of the
individuals infected has been derived~\cite{Newman2002}.  In particular we note
the use of a probability generating function~\cite{Volz2008,Miller_Note} to
derive a small number of nonlinear ODEs that describe the dynamics of a SIR
infection on a random heterogeneous network.  In most cases where networks are
used they are static, i.e.\ decided at one point in time and fixed that way
rather than changing over time, though there are several examples where this
assumption is not made~\cite{Kamp:2010,Volz_Myers,Miller_et_al}.  There are
also examples of networks on which the individuals have heterogeneous degree
and are always changing contacts~\cite{Pastor-Satorras:2001,May:2001}, although
there is some precedence for this modelling framework before the widespread use
of networks as a conceptual tool~\cite{May:1988}. Though the extremes of static
or extremely dynamic networks are obviously unrealistic, they are useful to
enable some analytic traction and also increase the ease with which we can
simulate. More recently though there have been attempts to include some
additional way of passing on the disease rather than just through the  defined
links of the network in an attempt to describe the fact that diseases can be
passed between members of the population that do not have more than one
interaction, i.e.\ a `global' infection term~\cite{Kiss2006}.

Observed epidemics are noisy and unpredictable, which motivates the use of
stochastic epidemic models~\cite{Andersson2000}. While some asymptotic results
for invasion and final size of stochastic epidemics on networks can be derived
in a discrete-time branching process framework, if one is interested in
transient dynamics then the natural model is an appropriate continuous time
Markov chain.  For an SIR-type epidemic model on an arbitrary graph with $N$
nodes, such a Markov chain would involve $3^N$ ODEs, which quickly becomes
computationally intractable.  The method proposed by Ball and
Neal~\cite{Ball2008}  involves creating a configuration model network at the same
time as the epidemic tree, and this stochastic process manifestly asymptotically
converges on
$2M$ ODEs in the deterministic large $N$
limit, where $M$ is the maximum number of contacts that the most well connected
individual on a network has. 
Unfortunately, this can still be very large
depending on the exact degree distribution and any attempt to cut down the
number of equations by ignoring individuals who have more than a given number
of neighbours, will inevitably ignore some of the most important individuals
for the dynamics. Fortunately, recent work~\cite{Decreusefond}  has shown
that a more sophisticated convergence proof leads to the smaller equation
set of~\cite{Volz2008}.  

The desire to model stochasticity without a massive increase in dimensionality
has led researchers to consider the diffusion limit.  This general approach to
stochastic processes is typically either attributed to van
Kampen~\cite{VanKampen1992} or Kurtz~\cite{Kurtz1970,Kurtz1971}, but the basic
idea is that provided that our population is sufficiently large, we can
approximate the Markov process that describes the epidemic by a deterministic
model together with appropriately scaled white noise processes that are defined
by the transition rates between the states of the Markov chain.  Such methods
have been used to derive a low dimensional model in which properties of the
noise in a stochastic epidemic model can be investigated
analytically~\cite{Alonso2007, Black2009}, by Ross~\cite{VRoss2006} to obtain
expressions for the mean and variance of a metapopulation model, and by Colizza
et al.\ \cite{Colizza} to model the effect of air travel on the spread of
epidemics in a large-scale network. These models are attractive since they have
the same dimensionality as the deterministic limit, but are actually
stochastic, and corrections to the approximation are $O(N^{-1})$.

In this paper we will apply the results of Kurtz to SIR-type epidemic dynamics
on a configuration model network, as was done for SIS dynamics on a regular
graph by Dangerfield et al.~\cite{Dangerfield2009}.  Using this we obtain a
four-dimensional set of stochastic ODEs, from which we derive an analytical
expression for the variance of the asymptotic early growth of an epidemic on a
network given its degree distribution. We provide an argument that our approach
is asymptotically exact, and simulate epidemics on various networks to confirm
the utility of our analytical results.

\section{Network model}

The use of networks as a generalisation from homogeneous mixing is becoming one
of the most widely used in epidemiological modelling. Contact between two
individuals of the population we are considering forms a link between them.
Once a link is established, the infection can be passed along it in either
direction.  Specifically what contact is represented depends on the disease,
i.e.\ it is different for a respiratory infection compared to a sexually
transmitted infection. These differences will affect how the contact network is
constructed, but we use a common modelling framework once the network is known.

The network can be described by a  symmetric $N\times{}N$ matrix $\mathbf{A}$  ,
which has binary entries, $A_{i,j}\in \{0,1\}$, where the entry $A_{i,j}$ will
be 1 if nodes $i$ and $j$ have a link between them and 0 otherwise.  We do not
allow self links in the network or multiple links between nodes.  We make the
simplifying assumptions that all links are of equal strength and that once a
link is made it will be there throughout the spread of the epidemic. 

The degree distribution of a network, $P(k)$, specifies what the probability is
that a node selected uniformly at random will have $k$ neighbours.
To construct an uncorrelated network with a given degree distribution, $P(k)$,
the configuration model is used~\cite{MolloyReed}. The method for this is as
follows:
\begin{itemize}
\item Each member of the population $i$ is given a number of ``half-links" or
``stubs", $k_i$, which is drawn from the degree distribution $P(k)$. Once this is done we can define
$d_k$ to be the proportion of nodes in the network that have $k$  neighbours.
\item Pairs of stubs are then picked uniformly at random and are then joined up
forming an undirected edge between the two corresponding vertices.
\end{itemize}
Later in the paper, when we simulate an epidemic on a network, this is the
method that is used to create a network with the degree distribution that we
require.  Asymptotically, the differences due to different ways of dealing with
repeated- and self-edges will be $O(N^{-1})$. Our choice for how to deal with
these in a finite system is to obtain a list of all the nodes which have
multiple connections between them and self-edges. One by one, the extra links
between nodes will be broken, say between node $i$ and $j$, and then a randomly
selected and connected pair of nodes will also be selected, say nodes $\imath'$
and $\jmath'$. We then connect $i$ and $\imath'$ together and $j$ and $\jmath'$
together, which leaves the degree distribution unaltered.  This is then
repeated until there are no repeated- and self-edges.

\section{Diffusion model}

\label{sec:diffmod}

\subsection{Model definition}

We assume a population of $N$ individuals on a configuration model network.
Individuals are stratified by their disease state $S$, $I$, or $R$, and their
degree on the network $k$. Individuals of type $S_k$ become $I_k$ at a rate
equal to the product of the transmission rate $\tau$ and their number of
infectious neighbours. Individuals of type $I_k$ become $R_k$ at a rate
$\gamma$. We are interested in the large $N$ regime, in which we write $[S_k]$
for the expected number of susceptibles of degree $k$, $[I_k]$ for the expected
number of infectious individuals of degree $k$, and $[AB]$ for the number of
  connected   pairs of individuals on the network where one is type
$A$ and the other is type $B$.  Omission of a subscript denotes implicit
summation, e.g. $[S] = \sum_k [S_k]$.

To derive pairwise equations, we require knowledge of the neighbourhood of each
node. This is due to the fact that when an infection or recovery takes place
the number of pairs of type $[AB]$ will be changed in a way which is dependent
on the neighbours that the node has. This is why it is not straightforward to
write down a low-dimensional form for this process as the population size
becomes large, since this requires the distribution of neighbours of each node.
We therefore make the following assumption, that we will justify below: The
distribution of neighbourhoods with $x$ susceptibles and $y$ infectious
individuals around a susceptible individual of degree $k$, is a multinomial
with probabilities that do not depend on $k$, i.e.
\begin{equation}
D^S_{x,y,k}=\binom{k}{x,y}%
(1-p_S-q_S)^{k-x-y}p_S^xq_S^y \text{ ,}
\label{eqn:mainassumption}
\end{equation}
where,
\begin{equation}
p_S = \frac{[S S]}{\sum_k k [S_k]} \mbox{ , } q_S = \frac{[S I]}{\sum_k k [S_k]} 
\label{eqn:connection_probabilities}\text{ .}
\end{equation}
Here the term $\binom{k}{x,y} = k!/x!y! $, is the multinomial coefficient.  We
have not detailed an assumption for the neighbourhoods of infected nodes here,
as this requires more care to be taken. This will be detailed later, and will
be denoted by $D_{x,k}^I$.  With these assumptions, it is possible to reduce
the state space of the Markov chain dramatically. In fact, a closed system is
obtained that is of dimension four.  The insight that allows this dimensional
reduction is from~\cite{Volz2008}, although our model uses pairwise notation
and is derived explicitly from applying the results of Kurtz to an underlying
stochastic process.  We note that $g(x) = \sum_k d_k x^k$ (where $d_k$, as
defined above, is the proportion of nodes with degree $k$) is the probability
generating function (pgf) of the degree distribution.  We also define and keep
track of the variable $\theta$ in our dynamical system. $\theta$ is defined to
be the proportion of degree one nodes that are still susceptible at time $t$.
Infection down each link is assumed to be independent, which implies that the
probability that a node of degree $k$ being susceptible at time $t$ will be
$\theta ^k$.  We can then write the number of degree $k$ nodes that are still
susceptible at time $t$ as $[S_k] = N d_k \theta(t)^k$. If there are no degree
one nodes in the population, then we can think of $\theta$ in terms of its
relationship with degree $k$ nodes. It is the $1/k$-th power of the probability
that a randomly chosen degree $k$ node is susceptible.

We also note that many important features of the network can be written simply
using the pgf. In particular: $N g(\theta) = N \sum_{k} d_k \theta^k = \sum_k
[S_k] = [S]  $, and $g'(1)  = \sum_k k d_k = \bar{n}$, where $\bar{n}$ is the
mean number of links per node.

To construct the deterministic limit of the epidemic on the large network
(called the fluid limit by some), we must consider the change in the variables
of our system due to an event.
 For the SIR model our events are infections of susceptibles and
recoveries of infecteds. As we are considering the epidemic on a network, to
fully describe the events, we are interested in how many neighbours someone has
who are infectious and susceptible. Keeping this in mind we therefore consider
two types of events. The first type of event is a susceptible of degree $k$,
who has $x$ susceptible neighbours and $y$ infected neighbours, becoming
infected, i.e. going from $S$ to $I$. The second is an infected individual of
degree $k$, who has $x$ susceptible neighbours, recovering from the infection,
i.e. going from $I$ to $R$. We denote these two events as $e_\tau(k,x,y)$ and
$e_\gamma(k,x)$ respectively, where $\tau$ and $\gamma$ are the rates of
infection and recovery, and write $\mathcal{E}$ for the set of such events. 

We then denote the rates of these events as $f_\tau(k,x,y)$ and
$f_\gamma(k,x)$. The rates are given by:
\begin{equation}
\begin{split} f_\tau(k,x,y) &= y \tau [S_k] D_{x,y,k}^S \text{ ,}\\
f_\gamma(k,x) &=  \gamma [I_k] D_{x,k}^I\text{ .} 
\end{split} 
\end{equation}  
We then sum the changes in our variables due to an event, multiplied by the
rate of that event, over all events to get the set of equations that we require. With this in mind, we have a state space $\mathbf{p} =
(\theta,[I],[SS],[SI])^\top$ obeying
\begin{equation}
\dot{\mathbf{p}} = \tau \sum_{k,x,y}  \Delta \mathbf{p}_{\tau}  \hspace{0.7mm} y \hspace{0.7mm} [S_k]\hspace{0.7mm} D^S_{x,y,k}  + \gamma \sum_{k,x} \Delta \mathbf{p}_{\gamma}   \hspace{0.7mm} [I_k] \hspace{0.7mm} D^I_{x,k} 
\text{ ,} \label{eqn:system_eq}
\end{equation}
where $\Delta \mathbf{p}_{\tau}$ and $\Delta \mathbf{p}_{\gamma}$ are the
changes to our variable under transmission and recovery respectively. We also
note that as we are not keeping track of the variable $[II]$, we will not be
concerned with the number of infected neighbours of an infective central node.
Therefore when we are concerned with making an assumption about the
neighbourhood of an infective, we will only be interested in the number of
susceptible neighbours that it has. To work out the change in $[SI]$ say due to
an infection, we have to consider the neighbourhood of the infected node. The
node in question was susceptible, becomes infected, meaning that the pairs
which were $[SS]$ pairs before infection, become $[SI]$ pairs and the $[SI]$
pairs will become $[II]$ pairs. If our node had $x$ susceptible neighbours and
$y$ infected neighbours, i.e. it formed $x$ $[SS]$ pairs and  $y$ $[SI]$ pairs,
then the change in $[SI]$ will be $x-y$. We can do similar calculations for the
other variables. We note that there is a subtlety in the calculations for
$\theta$ as we require the node which becomes infected to be of degree 1, and
it also depends on the number of nodes which are degree 1, given by $N d_1$.

Doing this we obtain the following expressions for $\Delta \mathbf{p}_{\tau}$
and $\Delta \mathbf{p}_{\gamma}$,
\begin{equation} 
\Delta \mathbf{p}_{\tau} = (-\delta_{k,1}/N d_1, 1, -2x,x-y)^\top \text{ , } \Delta \mathbf{p}_{\gamma} = (0,-1,0,-x)^\top \label{eqn:change_of_variables}\text{ , }
\end{equation}
which are the changes in the variables $\mathbf{p}$ from an infection and a
recovery respectively.  If we write down the exact but unclosed pairwise
equation for our system $\mathbf{p}$, then we get the following set of
equations:
\begin{equation}
\begin{split} \dot{\theta}  & = -  \tau  \frac{[SI]}{Ng'(\theta)}   \\
 \dot{[I]} & =  \tau [SI] - \gamma [I]   \\
\dot{[SS]} & =  - 2 \tau [SSI]  \\
\dot{[SI]} & = \tau( [SSI] -[ISI] -[SI]) - \gamma [SI] \text{ .}  \label{eqn:unclosed_system}
\end{split}
\end{equation}
We note that the expressions that we obtain for the recovery terms (terms
multiplied by $\gamma$) are the ones that we would require our assumption about
the neighbours of infecteds to calculate from \eqref{eqn:system_eq}. Due to the
fact that the equations concerning the recovery terms in
\eqref{eqn:unclosed_system} are exact and closed, we have two conditions that
we require $D_{x,k}^I$, to satisfy, \textit{viz.}
\begin{equation}
\sum_{k,x} [I_k] D_{x,k}^I = [I] \text{ ,}
\qquad
\sum_{k,x} x [I_k] D_{x,k}^I =[SI] \text{ .}
\label{eqn:conditions_for_infectious_neighbourhoods}
\end{equation}
Now when we evaluate \eqref{eqn:system_eq} using our main assumption
\eqref{eqn:mainassumption} we get the model of~\cite{Volz2008} in pairwise
notation. We note that the assumption of multinomially distributed neighbours
allows us to get a relatively simple set of equations, as the summation over
all events is equivalent to taking various moments of the distribution. We get
\begin{equation}
\begin{split}N \dot{\theta}  & = -  \tau  \frac{[SI]}{g'(\theta)}   \\
 \dot{[I]} & =  \tau [SI] - \gamma [I]   \\
\dot{[SS]} & =  - 2 \tau [SS][SI]  \frac{g''(\theta)}{N g'(\theta)^2}   \\
\dot{[SI]} & = \tau [SI]  \Big(\frac{g''(\theta)}{Ng'(\theta)^2} \Big([SS]-[SI] \Big) - 1 \Big) - \gamma [SI]
 \label{eqn:si} \text{ .}
 \end{split} \end{equation}
This can be thought of as a deterministic approximation to the underlying
stochastic evolution of the epidemic and is the exact limit of the  stochastic
process, which we describe in the next section, when $N \rightarrow \infty$
and~\eqref{eqn:mainassumption} holds. It is important to note that one cannot
simply start with~\eqref{eqn:si} and `add noise'.  Defining appropriate events
and rates, together with the explicit statement of~\eqref{eqn:mainassumption},
is essential. 

\subsection{Diffusion limit}

The work of Kurtz~\cite{Kurtz1970,Kurtz1971} also tells us that the noise
around the deterministic limit given above should be a Gaussian centred around
this limit with an appropriate density.  These results require
certain technical conditions to be satisfied in order to apply. In particular,
they require a family of right-continuous, temporally homogeneous jump Markov
processes with elements $\mathbf{X}_N$ indexed by an integer $N$. In our case,
$N$ is the number of nodes and the state of the Markov chain is $(X^S_{x,y,k},
X^I_{x,k})$ where $X^S_{x,y,k}$ is the number of degree-$k$ susceptible nodes
with $x$ susceptible and $y$ infectious neighbours, and $X^I_{x,k}$ is the
number of degree-$k$ infectious nodes with $x$ susceptible neighbours.
We seek a limit
\begin{equation}
N\rightarrow\infty \text{ ,} \qquad
X^S_{x,y,k} \rightarrow [S_k] D^S_{x,y,k} \text{ ,} \qquad
X^I_{x,k} \rightarrow [I_k] D^I_{x,k} \text{ ,}
\end{equation}
and then lump the equations derived to reduce the system dimensionality.  The
infinitesimal parameters for the rate of going from state $\mathbf{x}$ to state
$\mathbf{x}'$ also need to obey
\begin{equation}
Q^N_{\mathbf{x},\mathbf{x}'} = N F(N^{-1} \mathbf{x},\mathbf{x}') \text{ ,}
\end{equation}
for some function $F$.  If we can write the rates in this form, the general
results   give us the exact formula to work out what these noise terms
should be and the full stochastic equations in the diffusion limit are
\begin{equation}
\begin{split}
 \dot{\mathbf{p}} & =  \tau \sum_{k,x,y} \Delta\mathbf{p}_{\tau} \hspace{0.7mm} y \hspace{0.7mm}[S_k] \hspace{0.7mm}D^S_{x,y,k} \hspace{0.7mm}  +   \sum_{k,x,y} \Delta\mathbf{p}_{\tau} \sqrt{\tau  \hspace{0.7mm} y \hspace{0.7mm} [S_k] \hspace{0.7mm} D^S_{x,y,k}}  \hspace{0.7mm} \xi_\tau(t) \\ 
 & + \gamma \sum_{k,x}   \Delta\mathbf{p}_{\gamma} \hspace{0.7mm} [I_k] \hspace{0.7mm} D^I_{x,k}\hspace{0.7mm} +  \sum_{k,x}\Delta \mathbf{p}_{\gamma}\sqrt{\gamma \hspace{0.7mm}[I_k]\hspace{0.7mm} D^I_{x,k}} \hspace{0.7mm} \xi_\gamma(t)
\end{split} \text{ ,}
 \label{eqn:stochastic_eq}
\end{equation}
where $\xi_\tau(t)$ and $\xi_\gamma(t)$ are independent standard Gaussian noise
processes associated with transmission and recovery respectively.  We note that
there is no simple expression for the amplitudes of the noise processes, as the
square root prevents us from taking the moments of the multinomial distribution
and then expressing these in terms of the variables of our system, as we do for
the non-stochastic terms. Therefore we do not explicitly write the full
stochastic system, as the equations would be in a similar (but more complicated
looking) form to \eqref{eqn:stochastic_eq}.

We use methodology that is derived from~\cite{Kurtz1970,Kurtz1971} and set out
conveniently in~\cite{Dangerfield2009}, to analyse the variance in the
infection levels during the early growth of the epidemic. We define the early
growth period to be the time before there has been depletion to the pool of
susceptibles which is significant enough to affect the rate of growth in the
number of infecteds.  Our starting point is assuming that the early growth of
the infection is exponential, with rate $r$. That is,
\begin{equation}
 \left[I\right]  =  \tilde{I} \mbox{exp}(rt) \label{eqn:growth_assumption} \text{ , }
\end{equation}
where $\tilde{I}$ is a constant related to the prevalence of the infection as
the early asymptomatic behaviour commences.  Note that an additional
assumption in taking the diffusion limit is that this quantity should be
significantly larger than 1 but also significantly smaller than the total
population size $N$.  We then use this to work out the early behaviour
of the other variables,   \begin{equation}
\begin{split}
 \theta & =  1 - K_{\theta} \left[I\right] /N \text{ ,}\\
 \left[SS\right] & =  N g'(1) - K_{SS} \left[I\right] \text{ ,}\\
  \left[SI\right] & = K_{SI} \left[I\right]\text{ .} 
\end{split} 
\label{eqn:derived_behaviour}
\end{equation}
We now define the local covariance matrix associated with an event i.e.\
when a susceptible node becomes infected, or an infected node recovers. This is calculated as follows,
\begin{equation}
\mathbf{G}_{ij} = \sum_{e\in \mathcal{E}} f_e \Delta \mathbf{p}_{i,e} \Delta
\mathbf{p}_{j,e}  \label{eqn:G_Mat} \text{ ,}
\end{equation}
where $e \in \mathcal{E}$ means that $e$ is either an infection or recovery
event associated with a particular neighbourhood.  Using the early growth
Ansatz, specifically ignoring terms $O([I]^2)$, we find that this can be
written as  $\mathbf{\hat{G}}[I](t)$, where $ \mathbf{\hat{G}}$ is constant.
We can now use the following equation from~\cite{Dangerfield2009}, which is
again derived from the theoretical work of Kurtz~\cite{Kurtz1970,Kurtz1971},
\begin{equation}
 r \mbox{\boldmath$\sigma$}^2 -\mathbf{B}  \mbox{\boldmath$\sigma$}^2 - \mbox{\boldmath$\sigma$}^2 \mathbf{B}^{T} = [ \mathbf{\hat{G}} \mbox{exp}(r t)
  - \mbox{exp}(\mathbf{B} t)\mathbf{\hat{G}}\mbox{exp}(\mathbf{B} t)^T ] \tilde{I} \label{eqn:matrices} \text{ ,}
\end{equation}
where  $\mbox{\boldmath$\sigma$}^2$ is the time dependent covariance matrix of our state variables $\mathbf{p}$,
$r$ is the early exponential growth rate, $\mathbf{B}$ is the Jacobian of the
fluid limit of our system \eqref{eqn:si}, evaluated using the early growth
Ansatz \eqref{eqn:growth_assumption} and \eqref{eqn:derived_behaviour}, and
$\mathbf{\hat{G}}$ is given above.  The
aim is to find an expression for $\mbox{\boldmath$\sigma$}^2$, as this will
give us the variance of the epidemic during the early growth phase. 

We now give specific details of how the various matrices in the above equation can be calculated. 
To calculate $\mathbf{\hat{G}}$, we calculate the matrix $\mathbf{G}$ and then divide by $[I](t)$. To do this we
substitute in the changes in variables $\Delta \mathbf{p}$, which are given above by
\eqref{eqn:change_of_variables}. This means that we can write
$\mathbf{G}$ as follows,
\begin{equation}
\mathbf{G} = \tau \sum_{k,x,y} y [S_k] D^S_{x,y,k} \mathbf{F}_{\tau}
+\gamma \sum_{k,x} [I_k] D^I_{x,k} \mathbf{F}_{\gamma} \text{ ,}
\end{equation}
where 
\begin{equation}\mathbf{F}_{\tau} =\left(
\begin{array}{cccc}
 \frac{\delta _{1,k}}{N^2 d_1^2} & -\frac{\delta _{1,k}}{N d_1} & \frac{2 x \delta _{1,k}}{N d_1} & \frac{(y-x) \delta _{1,k}}{N d_1} \\
 -\frac{\delta _{1,k}}{N d_1} & 1 & -2 x & x-y \\
 \frac{2 x \delta _{1,k}}{N d_1} & -2 x & 4 x^2 & 2 x (y-x) \\
 \frac{(y-x) \delta _{1,k}}{N d_1} & x-y & 2 x (y-x) & (x-y)^2 
\end{array}
\right) \text{ ,} \quad
\mathbf{F}_{\gamma}=\left(
\begin{array}{cccc}
 0 & 0 & 0 & 0 \\
 0 & 1 & 0 & x \\
 0 & 0 & 0 & 0 \\
 0 & x & 0 & x^2
\end{array}
\right) \text{ ,}
\end{equation}
are the outer products of the change in each variable vectors $\Delta
\mathbf{p}_\tau$ and $\Delta \mathbf{p}_\gamma$. 

Let us consider how this calculation is done in practice by working out the
entry $\mathbf{G}_{[I],[SS]}$. We use the (2,3) entry of
$\mathbf{F}_{\tau} = -2x $ and $\mathbf{F}_{\gamma} = 0$, which gives us that
$\mathbf{G}_{[I],[SS]} = - 2 \tau \sum_{k,x,y} [S_k] \mbox{ } xy \mbox{ }
D_{x,y,k}^S$. We can separate the sum over $k$ from the sum over $x$ and $y$,
and we can use the fact that the (1,1) moment of a multinomial distribution
with variables $k,p$ and $q$ is given by $k(k-1) p q$. Therefore
$\mathbf{G}_{[I],[SS]} = - 2 \tau \sum_{k,x,y} [S_k]\mbox{ } k
(k-1)\mbox{ }p_S q_S$. We can then use our assumption about the probabilities
of connecting to susceptibles or infecteds from
\eqref{eqn:connection_probabilities} and sum over the indices. We get:
\begin{equation}
 \mathbf{G}_{[I],[SS]} = - 2  \tau[SS][SI]\frac{g''(\theta)}{N g'(\theta)^2}
\text{ .}
\end{equation}
When we have calculated all the terms for the $ \mathbf{G}$ matrix we
will input \eqref{eqn:growth_assumption} and \eqref{eqn:derived_behaviour} to
obtain the correct matrix for the early growth period.  We do this for
evaluate this for $\mathbf{G}_{[I],[SS]}$  linearising it with respect to
$[I]/N$, i.e. any term which is $O(([I]/N)^2)$ will be ignored. We then obtain
the following expression for $\mathbf{G}_{[I],[SS]}$:

\begin{equation}
\mathbf{G}_{[I],[SS]}     =   -2 \tau [I]  \frac{g''
   \left(g'' - g'\right)  }{(g')^2} +O([I]/N)^2\text{ ,}
\end{equation}
where we have used the fact that $g'(\theta)$ and $g''(\theta)$ will become
$g'(1)$ and $g''(1)$ with the use of the early growth Ansatz, and then writing this in terms
of $g$, where we define $g^{(n)} \equiv g^{(n)}(1)$. This now gives us that,
\begin{equation}
\mathbf{\hat{G}}_{[I],[SS]}     =   -2 \tau  \frac{g''
   \left(g'' - g'\right)  }{(g')^2} \text{ .}
\end{equation}
We show all the entries of $\mathbf{\hat{G}}$ in Appendix B.

The $\mathbf{B}$ matrix from \eqref{eqn:matrices} as stated above is the
Jacobian matrix of the deterministic limit of the system, evaluated using
\eqref{eqn:growth_assumption} and \eqref{eqn:derived_behaviour} and is
therefore given by
\begin{equation}
\mathbf{B}=\left(
\begin{array}{cccc}
 0 & 0 & 0 & -\frac{\tau}{Ng'(1)}  \\
 0 & -\gamma  & 0 & \tau  \\
 0 & 0 & 0 & -\frac{\tau g''(1) }{g'(1)} \\
 0 & 0 & 0 & r
\end{array}
\right) \text{ .}
\end{equation}
Now we have calculated $\mathbf{\hat{G}}$ and $\mathbf{B}$ we can solve
\eqref{eqn:matrices} for $\mbox{\boldmath$\sigma$}^2$. 

\subsection{Neighbourhoods around an infected node}

\label{sec:ninf}

The discussion above has so far not dealt with the neighbourhood of an
infective. This only becomes an important consideration for the bottom right
term in $\mathbf{F}_{\gamma}$, where we have to work out the square of the
change in $[SI]$ pairs due to recovery of an infectious node;  in
particular we are interested in calculation of the quantity
\begin{equation}
\chi := \sum_{k,x} x^2 [I_k] D_{x,k}^I \label{eqn:second_moment} \text{ ,} 
\end{equation}
and so the task is to determine the number of infectives of degree $k$, and the
second moment of the distribution of the number of susceptibles around such
infectives. Now, the rate at which infectives of degree $k$ are produced is
given by
\begin{equation}
-\frac{\mathrm{d}}{\mathrm{d}t}[S_k] = -N d_k \frac{\mathrm{d}}{\mathrm{d}t}
\theta^k = -N d_k k  \dot{\theta} \theta^{k-1} \text{ .}
\end{equation}
We also know that the probability that an infective of age $a$ (a node which
was infected length of time $a$ ago) is still infective is given by $e^{-\gamma
a}$.  We can therefore work out $[I_k]$ at a given time $t$ by evaluating the
following integral:
\begin{equation}
[I_k] = -N d_k k \int_0^t \dot{\theta}(t-a) (\theta(t-a))^{k-1}
e^{-\gamma a}\; \mathrm{d}a \text{ .}
\end{equation}
Now consider the neighbourhood around an such an infective. Ignoring the very
first cases, every infectious individual must have been infected by someone,
leaving $k-1$ individuals who are potentially susceptible. If the infection of
the central node happened a time $a$ ago, then each of the $k-1$ potentially
susceptible neighbours has an independent probability $e^{-\tau a}$ of avoiding
infection from that central node, and in the event that central infection is
avoided and the neighbouring node is of degree $l$ a probability of
$\theta^{l-1}$ of avoiding infection from any other source. Summing over $l$
appropriately weighted then gives the general expression
\begin{equation}
[I_k] D_{x,k}^I = -N d_k k \int_0^t \dot{\theta}(t-a) (\theta(t-a))^{k-1}
e^{-\gamma a}\;
 \mathrm{Bin}\bigg(x\; \bigg|\; k-1,\; \frac{g'(\theta(t))}{g'(1)} e^{-\tau a}\bigg) \;
\mathrm{d}a \text{ ,}
\end{equation}
where $\mathrm{Bin}(x|k,\pi)$ is the binomial probability mass function,
representing the probability that $x$ of $k$ trials with independent
probability of success $\pi$ will be successful. Since we are interested in
early asymptotic behaviour, we then linearise this expression making use of
the expressions
\begin{equation}
\theta \rightarrow 1 \text{ ,}\qquad
\dot{\theta} = -r \frac{K_{\theta}}{N} [I] \text{ .}
\end{equation}
This gives asymptotic results
\begin{equation}
\begin{aligned}
\left[I_k\right] D_{x,k}^I & \rightarrow [I]  \frac{r+\gamma}{g'(1)}
\int_0^t e^{-(r+\gamma)a}\; \mathrm{Bin}(x|k-1,e^{-\tau a})\; {\rm d}a \\
\Rightarrow \chi & \rightarrow [I]  \frac{r+\gamma}{g'}\bigg(\frac{g''}%
{\tau +r +
\gamma} + \frac{g'''}{2 \tau +r + \gamma} \bigg) \text{ .}
\end{aligned}
\end{equation}
We have now calculated the entirety of the $\mathbf{G}$ matrix and can express it in the form $\mathbf{\hat{G}}[I]$, where
$\mathbf{\hat{G}}$ is constant, as is required to use the results of Kurtz.

\subsection{Rate of convergence}

We now consider the rate at which convergence to our model will happen. Our
approach here is heuristic rather than formal, and is based on the fundamental
contributions by Ball and Neal~\cite{Ball2008}, and more recently Decreusefond
et al.~\cite{Decreusefond} We start by redefining the network and epidemic
processes in ways that are less useful for practical calculation than our
initial definitions, but which make the rate of convergence more clear.  In
each case, we start with $N$ individuals indexed $i,j,\ldots$

In a \textit{configuration model process}, we let individual $i$ have $K_i$
stubs, and start the network ajacency matrix $A_{i,j}(t=0)=0, \forall i,j$.
Then the process is defined to take
\begin{equation}
(K_i, K_j, A_{i,j}, A_{j,i}) \rightarrow 
(K_i-1, K_j-1, A_{i,j}+1, A_{j,i}+1) \text{ at rate} 
\propto K_i K_j \text{ .}
\end{equation}
Running this process until the absorbing state $K_i = 0, \forall i$ gives the
adjacency matrix of a configuration model network, although depending on what
one decides about repeated- or self-edges, different corrections of $O(N^{-1})$
may arise~\cite{Durrett:2007}.

In the \textit{epidemic process}, individuals have non-independent random
variables for their state $X_i$, that can take values $S$, $I$ or $R$. Then
\begin{equation}
(S_i,I_j) \rightarrow 
(I_i,I_j) \text{ at rate } 
\tau A_{ij} \text{ ,}
\end{equation}
and infectious nodes recover at rate $\gamma$. The fundamental insight from
Ball and Neal~\cite{Ball2008} is that these two processes can be combined.  In
this construction, every node in the network is given a number of half-links,
which are then paired up as the epidemic progresses to form contacts between
individuals in the following two ways:
\begin{itemize}
\item An infected with $l$ remaining half-links makes contacts at rate $\tau
l$; if it links to a susceptible then the infection will be transmitted with
probability 1.
\item When an infected recovers (which happens at rate $\gamma$) all of its
remaining half-links will be independently paired off with remaining half-links
in the population.
\end{itemize} 
The question of what a susceptible neighbourhood looks like at a given time in
this picture is answered by halting the epidemic process, and running the
configuration model process to completion, then counting the neihbours of each
susceptible. As noted by Decreusefond et Al.~\cite{Decreusefond} (who use a
slightly different construction but the same basic idea) at finite $N$ this
gives each susceptible a multivariate hypergeometric neighbourhood, as they
sample without replacement from the population. As $N$ becomes large, this
tends to a multinomial distribution with corrections $O(N^{-1})$.

A similar argument can be made for the neighbourhood around an infectious
individual, in terms of its links that remain unpaired. In the absence of
susceptible depletion, it is then possible to make a stochastic version of
the deterministic argument about the neighbourhood around an infective in 
\S{}\ref{sec:ninf} above.

Since the diffusion limit deals with terms of $O(N^{-1/2})$ and ignores terms
of $O(N^{-1})$, we are therefore justified in making the multinomial assumption
for the neighourhood around a susceptible.  We note that it therefore implies
the exactness of several other deterministic approaches in the absence of
clustering, such as the pairwise closure techniques from \cite{Keeling1999} and
the maximum entropy moment closure method of Rogers~\cite{Rogers}.

\subsection{Early growth variance}

As described above we can solve \eqref{eqn:matrices} for
$\mbox{\boldmath$\sigma$}^2$.  We did this through computer algebra, due to the
complexity of expressions involved. The variance of the number of infecteds
during early growth is shown in full in Appendix A.  This can be simplified in
the certain regimes. If we define the time at which the epidemic grows at the
rate predicted in \cite{Diekmann2000} as $t_\mathrm{early}$, and the time at
which  the depletion of susceptibles affects the growth rate and we leave the
early growth phase at $t_\mathrm{depleted}$, then when the current time
satisfies $t_\mathrm{early} \ll t \ll t_\mathrm{depleted}$, we can simplify the
expression in Appendix A. We note that this regime will exist in an infinite
size network if the initial amount of infection in the network $\tilde{I}$ is
sufficiently small.  It is dominated by a single term which involves the mean,
variance and skew of the networks degree distribution along with the recovery
and transmission rates of the infection. In this limit, the mean and variance
of prevalence obey  
\begin{equation}
\begin{split}
\text{Mean}(I) & \rightarrow \tilde{I}e^{rt} \text{ ,} \text{\ \ for} \quad
r = \tau \left(\frac{g''}{g'} -1\right) - \gamma
\text{ ,}\\
\frac{\text{Var}(I)}{\text{Mean}(I)^2}  & \rightarrow 
 \frac{ g'^2 (g'''+(\tau +1) (g''+g'))}{(g'^2-g''^2) ((\gamma +\tau ) g'-\tau  g'')} 
  \text{ .}
\label{eqn:t_inf_var}
\end{split}
\end{equation}
 where $\tilde{I}$ is a constant related to the prevalence of infection
as the early asymptotic behaviour commences at $t_\mathrm{early}$,  and
$g^{(n)} \equiv g^{(n)}(1)$.

\begin{figure}[H]

 \begin{center}
\includegraphics[width=0.75\textwidth]{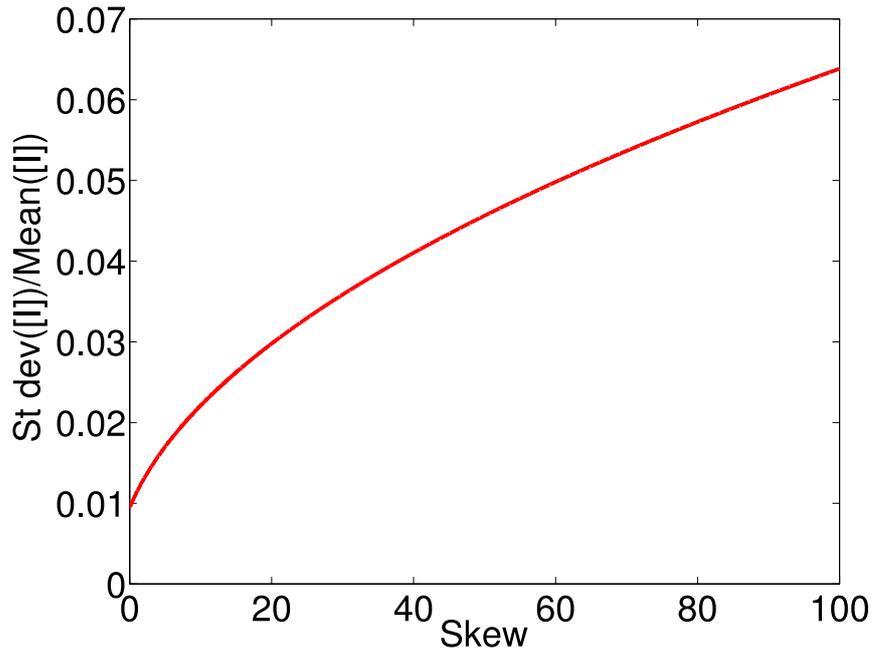} 
\caption{Early asymptotic dependence of the standard deviation of infection
prevalence, divided by infection prevalence, on the skew $\Gamma$ of the
network degree distribution. The curve is plotted from
equation~\eqref{eqn:t_inf_var} in the main text. Parameter values are: mean
degree $\approx 5.4$; variance in degree $\approx 67.2$; transmission rate
$\tau = 0.0308$; recovery rate $\gamma = 0.1$.  Skewness of degree is varied
between realistic values (0 and 100) to see how the variance of the asymptotic
early prevalence of infection is affected.  We see that as the skewness is
increased we get a higher variability of prevalence. This is as expected, since
the higher the skew, the more neighbours the most connected individuals of the
population have, reducing the predictability of the epidemic due to chance
events amongst this small but epidemiologically important group.} 
\label{fig:skew_t_inf}
 \end{center}

\end{figure}

Figure~\ref{fig:skew_t_inf} shows how the variance in prevalence changes as
skew in degree increases. Seeing that as the skew increases we get an increase
in the variance of the epidemic during early growth, we would therefore see
that if we had a network whose degree distribution is a power law, would show
greater variation than a negative binomially distributed network with the same
mean and variance would show. We think of this increase in variation being
caused by the members of the population who are very well connected in the
network. These people have been called super-spreaders in the past. If the
disease reaches these people in the early growth phase of the epidemic then we
can expect a rapid increase in disease prevalence, as there will be many $[SI]$
pairs through which the disease can be passed, whilst if they do not get it in
this early phase, there will not be this rapid increase, which generates the
large variance in this stage of the epidemic. We note that for the $\tilde{I}$
term in \eqref{eqn:t_inf_var} we have no analytical traction on what this
should be for a given epidemic. If we wish to compare this result to
simulation, we will fit the value of $\tilde{I}$ so that the analytical
prediction and the simulated results agree at a given point. As this is simply
multiplied by the other terms, fitting it simply scales the prediction up or
down rather than fine tuning the prediction itself.

\section{Comparison with simulation}

We have also used simulation to test our conclusion that if we fix the mean and
variance of the degree distribution, but have different skew values, then the
network which has the higher value of skew will also exhibit a higher variance
of infecteds during the early growth period. We wish to see whether we get a
match between the analytical result that we have obtained and the simulated
result. There are several difficulties to achieving this that we wish to note.
First, the networks are extremely large in the case of the analytical results
and we do not know how exactly fast the convergence to the answer should be
(although the corrections are $O(N^{-1})$ we do not know what prefactors
multiply this term). We have run our simulations on networks of size $10^5$,
representing a small city, which in the end seemed sufficient. 

Secondly, the analytical results only work for the early growth phase of the
infection, which can be defined as the time when the pool of susceptibles is
not significantly depleted. Whilst in an arbitrarily large network this can
last for an arbitrarily long time, in a finite network, this is not true and
will vary depending on the degree distribution of the network. This means that
if we want to compare the early growth variance of two networks, we may have a
very brief window in which we can actually do it. This is exacerbated by the
fact that we will have to allow for a period at the very beginning of the
epidemic to be discounted, as we want the system to approach some sort of early
asymptotic behaviour unaffected by the random selection of initial infected.

Finally, there were also several assumptions made for the analytical system
such as how the number of triples in the network can be approximated by doubles
and the pgf $g(x)$.  With the finite network that we construct there is no
guarantee that this will be accurate; again, convergence is $O(N^{-1})$ but we
do not have knowledge of the prefactors.

We implemented the Gillespie algorithm~\cite{Gillespie1977} to simulate the
time evolution of the epidemic, on networks of size $10^5$.  To obtain the
correct mean early growth behaviour, we allow each simulation to achieve a
certain number of infections ($10^2$) and then we set the simulation time to
zero and let the epidemic progress from there.  By allowing this initial amount
of infection we will be able to absorb the initial conditions that we choose
and get to the average behaviour of the system. In the system of size $10^5$,
allowing $10^2$ infections is also small enough that the susceptibles will not
be depleted significantly enough to affect the rate of growth of the epidemic.
We ran $10^3$ simulations on two networks with the same mean and variance but
with different skewness. We note that the same networks were used for each of
the $10^3$ simulations.

\begin{figure}[H]
\centering
\subfloat[]{\includegraphics[width = 0.75\textwidth]{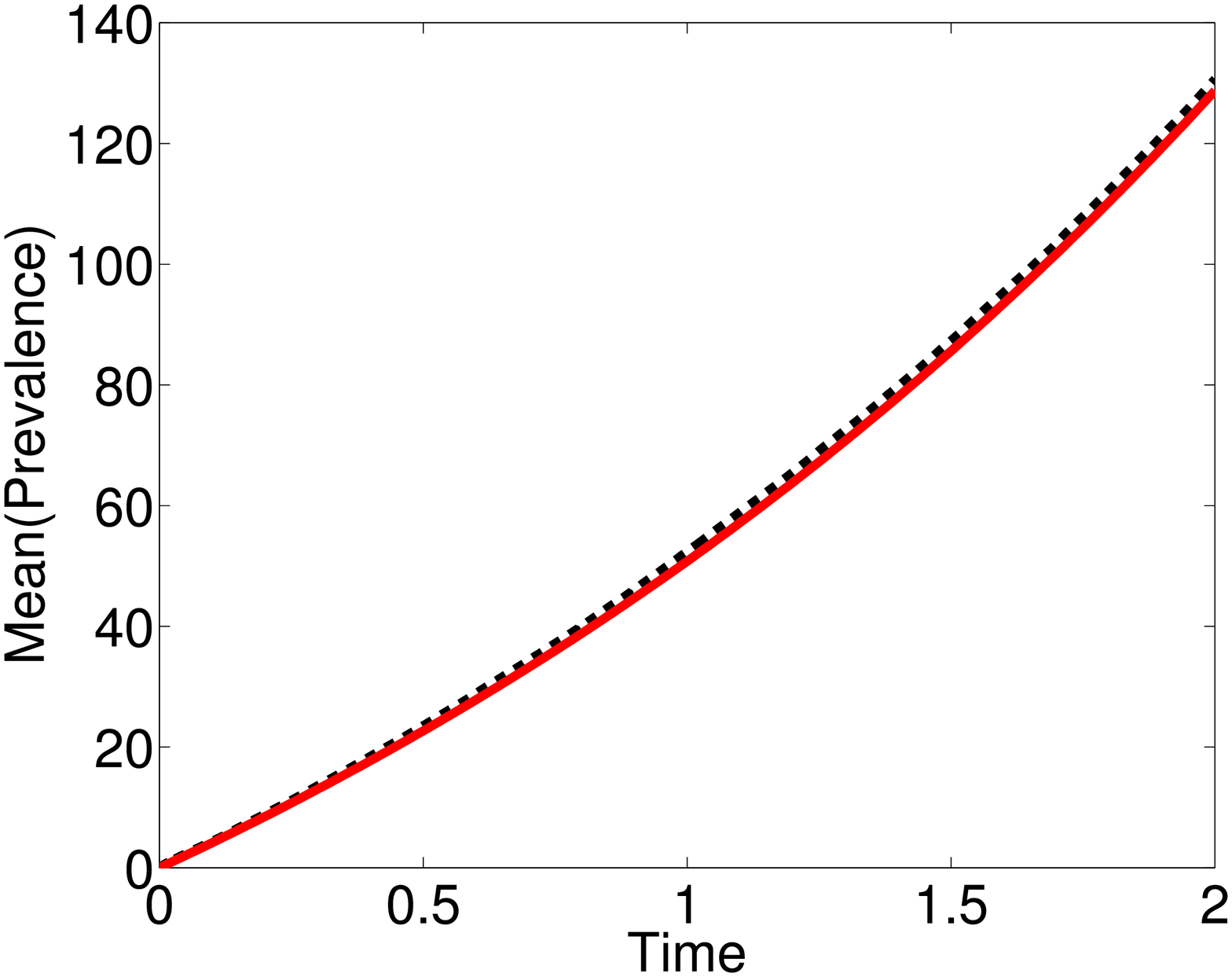}}\\
\subfloat[]{\includegraphics[width = 0.75\textwidth]{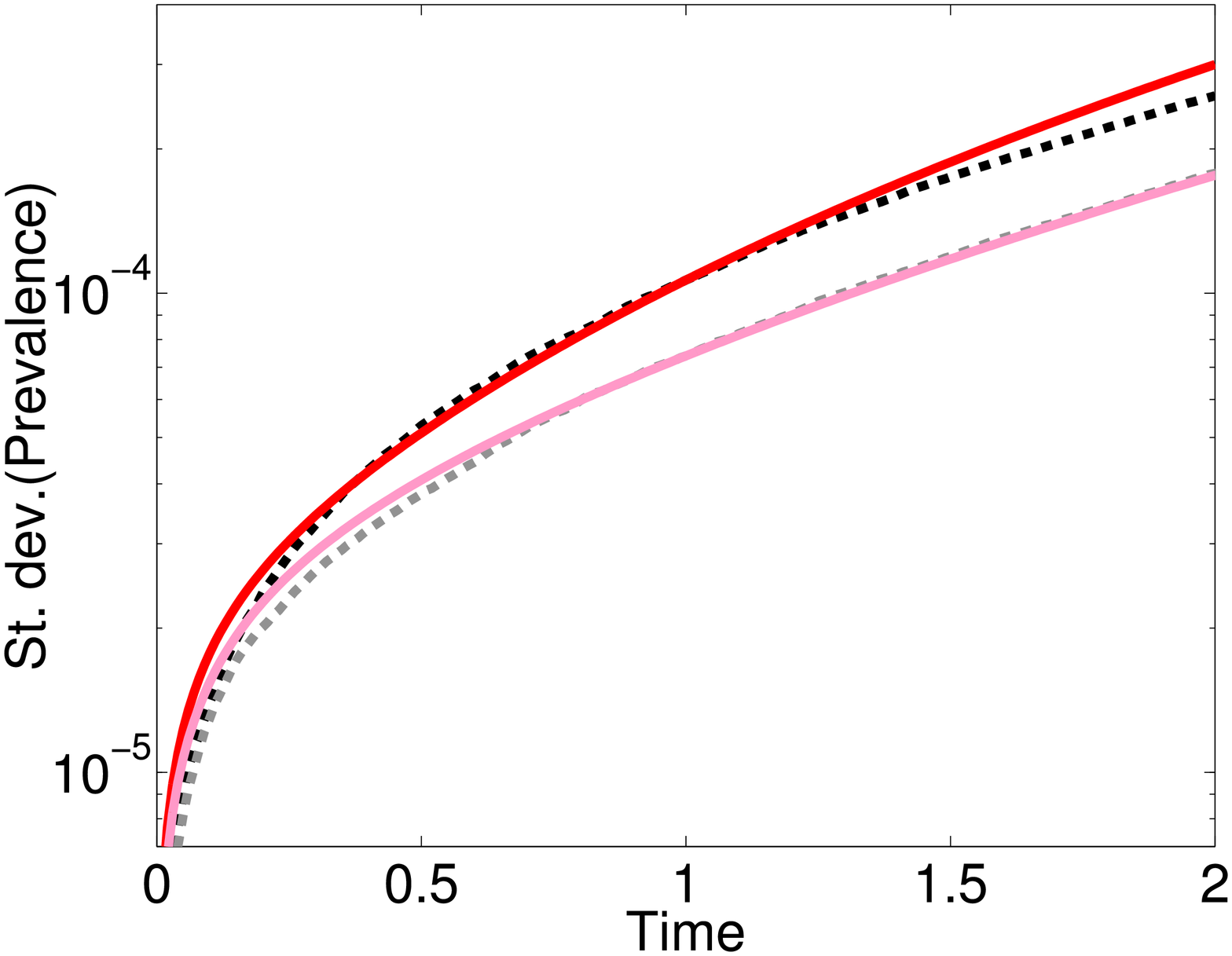}}
\caption{\label{fig:compare_sim_to_theory} Comparison of simulated results to
analytical predictions. Dashed lines are simulations and full lines are
analytical predictions. Simulations are on two different networks, which have
the same mean and variance for their degree distribution (mean $\approx 5.4$
and variance $\approx 67.2$) but different skewness: 24.3 for red / black
lines and 6.7 for the pink / grey lines. Transmission and recovery rates are
$\tau = 0.0308$ and $\gamma = 0.1$ respectively.  (a) shows a period of time at
which we have agreement in the growth of the number of infecteds between the
two networks that is strongly in agreement with the theoretical prediction from
\cite{Diekmann2000}. (b) is also taken for this time and we can see that the
theoretical prediction that we have described previously in this paper deviates
slightly from simulation, with this most pronounced early on when $O(N^{-1})$
corrections should be most significant.}
\end{figure}

Figure \ref{fig:compare_sim_to_theory} shows the result of these simulations
compared with the prediction given by \eqref{eqn:Variance_I} in Appendix A.  We
can see that as predicted the network with the higher skew exhibits more
variance in number of infecteds in the early growth stage and also we see that
the analytical prediction for what the variance should be, is a good fit for
what the simulations show. The early time disagreement is as we would expect,
since this is when the $O(N^{-1})$ corrections should be most significant.

\section{Discussion}

We have considered the spread of SIR-type infections on networks of
heterogenous degree. Using the assumption that the neighbourhood around a node
will have a distribution of susceptibles and infecteds which is multinomial
\eqref{eqn:mainassumption}, we have derived a low dimension deterministic
approximation \eqref{eqn:si} to the stochastic dynamics of the infection
which we can write down precisely in the form given by equation
\eqref{eqn:stochastic_eq}. A heuristic argument has been given to justify this
assumption in the large $N$ regime.

Computer algebra was then used to calculate the covariance matrix of our model
using an equation derived by Kurtz. This was then used to give us the variance
of the early growth of the epidemic on the network and we extracted the
dominant term in the $t_\mathrm{early} \ll t \ll t_\mathrm{depleted}$ regime
which is given by \eqref{eqn:t_inf_var}. These analytic results were derived
for networks of extremely large size.

By comparing the solution that was derived analytically with simulations, we
have been able to demonstrate that with a network of size $10^5$ we can get a
strong agreement between the two as can be seen in
Figure~\ref{fig:compare_sim_to_theory}.

We have shown that as the degree skewness of the network increases, so does the
variance of the number of infecteds during the early growth. An implication of
this result is that in the situation of an outbreak of a disease, if we are
able to target the very well connected people in the network, then we can (as
well as has been previously established) decrease the impact of the disease
efficiently -- but such an intervention would also mitigate the `reasonable
worst case scenario' in how many people will become infected by reducing the
variability of the epidemic.

Another potential application of our results is to improve the estimation of
epidemiological parameters during early growth of an epidemic. The possibility
of estimating the variability in prevalence, together with a relationship
between such variability and underlying model parameters, could enhance
statistical work on epidemic prevalence curves.

In conclusion, we have shown how a low-dimensional system of stochastic
differential equations can be used to describe the diffusion limit of a
stochastic epidemic on a heterogeneous network, and have drawn out
epidemiological conclusions from this model.

\section*{Acknowledgements}

The authors would like to acknowledge the financial support received from
EPSRC, the editor and two anonymous reviews for their helpful comments, in
particular relating to the analyses presented in \S{}\ref{sec:ninf} and
Appendix \ref{sec:bp}.

\bibliographystyle{abbrv}

\begin{thebibliography}{10}

\bibitem{Alonso2007}
D.~Alonso, A.~J. McKane, and M.~Pascual.
\newblock {Stochastic amplification in epidemics}.
\newblock {\em Journal of the Royal Society Interface}, 4(14):575--582, 2007.

\bibitem{Anderson&May1992}
R.~M. Anderson and R.~M. May.
\newblock {Infectious Diseases of Humans; Dynamics and Control}.
\newblock {\em Epidemiology and Infection}, 108(1), 1992.

\bibitem{Andersson2000}
H.~Andersson and T.~Britton.
\newblock {\em Stochastic Epidemic Models and Their Statistical Analysis},
  volume 151 of {\em Springer Lectures Notes in Statistics}.
\newblock Springer, Berlin, 2000.

\bibitem{Baguelin:2010}
M.~Baguelin, A.~J.~V. Hoek, M.~Jit, S.~Flasche, P.~J. White, and W.~J. Edmunds.
\newblock Vaccination against pandemic influenza {{A/H1N1v in England}}: A
  real-time economic evaluation.
\newblock {\em Vaccine}, 28(12):2370--2384, Mar 2010.

\bibitem{Ball2008}
F.~Ball and P.~Neal.
\newblock {Network epidemic models with two levels of mixing.}
\newblock {\em Mathematical Biosciences}, 212(1):69--87, Mar. 2008.

\bibitem{Bansal2007}
S.~Bansal, B.~T. Grenfell, and L.~A. Meyers.
\newblock {When individual behaviour matters: homogeneous and network models in
  epidemiology.}
\newblock {\em Journal of the Royal Society Interface}, 4(16):879--91, Oct.
  2007.

\bibitem{Black2009}
A.~Black, A.~McKane, A.~Nunes, and A.~Parisi.
\newblock {Stochastic fluctuations in the susceptible-infective-recovered model
  with distributed infectious periods}.
\newblock {\em Physical Review E}, 80(2):21922, 2009.

\bibitem{Colizza}
V.~Colizza, A.~Barrat, M.~Barth\'{e}lemy, and A.~Vespignani.
\newblock {The role of the airline transportation network in the prediction and
  predictability of global epidemics}.
\newblock {\em PNAS}, 103(14), 2006.

\bibitem{Dangerfield2009}
C.~E. Dangerfield, J.~V. Ross, and M.~J. Keeling.
\newblock {Integrating stochasticity and network structure into an epidemic
  model.}
\newblock {\em Journal of the Royal Society Interface}, 6(38):761--74, Sept.
  2009.

\bibitem{Danon:2011}
L.~Danon, A.~P. Ford, T.~House, C.~P. Jewell, M.~J. Keeling, G.~O. Roberts,
  J.~V. Ross, and M.~C. Vernon.
\newblock Networks and the epidemiology of infectious disease.
\newblock {\em Interdisciplinary Perspectives on Infectious Diseases},
  2011:1--28, Jan 2011.

\bibitem{Decreusefond}
L.~Decreusefond, J.-S. Dhersin, P.~Moyal, and V.~C. Tran.
\newblock Large graph limit for a {{SIR}} process in random network with
  heterogeneous connectivity.
\newblock {\em Annals of Applied probability}, 22(2):541--575, 2012.

\bibitem{Diekmann2000}
O.~Diekmann and J.~A. {P Heesterbeek}.
\newblock {\em {Mathematical epidemiology of infectious diseases}}.
\newblock John Wiley \& Sons Ltd., 2000.

\bibitem{Durrett:2007}
R.~Durrett.
\newblock {\em Random Graph Dynamics}.
\newblock Cambridge University Press, 2007.

\bibitem{Eames:2002}
K.~T.~D. Eames and M.~J. Keeling.
\newblock Modeling dynamic and network heterogeneities in the spread of
  sexually transmitted diseases.
\newblock {\em PNAS}, 99(20):13330--13335, Jan 2002.

\bibitem{Ferguson:2003}
N.~Ferguson, M.~Keeling, W.~Edmunds, R.~Gant, B.~Grenfell, R.~Amderson, and
  S.~Leach.
\newblock Planning for smallpox outbreaks.
\newblock {\em Nature}, 425(6959):681--685, Jan 2003.

\bibitem{Gillespie1977}
D.~T. Gillespie.
\newblock {Exact stochastic simulation of coupled chemical reactions}.
\newblock {\em The Journal of Physical Chemistry}, 81(25):2340--2361, Dec.
  1977.

\bibitem{Kamp:2010}
C.~Kamp.
\newblock Untangling the interplay between epidemic spread and transmission
  network dynamics.
\newblock {\em PLoS Computational Biology}, 6(11):e1000984, Nov 2010.

\bibitem{Keeling1999}
M.~J. Keeling.
\newblock {The effects of local spatial structure on epidemiological
  invasions.}
\newblock {\em Proceedings of the Royal Society B}, 266(1421):859--67, Apr.
  1999.

\bibitem{Keeling&Rohani2008}
M.~J. Keeling and P.~Rohani.
\newblock {\em {Modeling Infectious Diseases}}.
\newblock Princeton University Press, 2008.

\bibitem{Kiss2006}
I.~Z. Kiss, D.~M. Green, and R.~R. Kao.
\newblock {The effect of contact heterogeneity and multiple routes of
  transmission on final epidemic size.}
\newblock {\em Mathematical Biosciences}, 203(1):124--36, Sept. 2006.

\bibitem{Kurtz1970}
T.~G. Kurtz.
\newblock {Solutions of ordinary differential equations as limits of pure jump
  Markov processes}.
\newblock {\em Journal of Applied Probability}, 7(1):49--58, 1970.

\bibitem{Kurtz1971}
T.~G. Kurtz.
\newblock {Limit Theorems for Sequences of Jump Markov Processes Approximating
  Ordinary Differential Processes}.
\newblock {\em Journal of Applied Probability}, 8(2):344--356, 1971.

\bibitem{Lindquist2011}
J.~Lindquist, J.~Ma, P.~{van den Driessche}, and F.~H. Willeboordse.
\newblock {Effective degree network disease models.}
\newblock {\em Journal of Mathematical Biology}, 62(2):143--164, 2011.

\bibitem{May:1988}
R.~M. May and R.~M. Anderson.
\newblock The transmission dynamics of human immunodeficiency virus {{(HIV)}}.
\newblock {\em Philosophical Transactions of the Royal Society of London Series
  B}, 321(1207):565--607, 1988.

\bibitem{May:2001}
R.~M. May and A.~L. Lloyd.
\newblock Infection dynamics on scale-free networks.
\newblock {\em Physical Review E}, 64(066112), 2001.

\bibitem{Miller_Note}
J.~C. Miller.
\newblock A note on a paper by {{Erik Volz: SIR}} dynamics in random networks.
\newblock {\em J Math Biol.}, 62(3):349--358, Mar 2010.

\bibitem{Miller_et_al}
J.~C. Miller, A.~C. Slim, and E.~Volz.
\newblock Edge-based compartmental modelling for infectious disease spread.
\newblock {\em Journal of the Royal Society Interface}, 9(70):890--906, May
  2012.

\bibitem{MolloyReed}
M.~Molloy and B.~Reed.
\newblock A critical point for random graphs with a given degree sequence.
\newblock {\em Random Structures \& Algorithms}, 6(2/3):161--179, 1995.

\bibitem{Mossong:2008}
J.~Mossong, N.~Hens, M.~Jit, P.~Beutels, K.~Auranen, R.~Mikolajczyk,
  M.~Massari, S.~Salmaso, G.~S. Tomba, J.~Wallinga, J.~Heijne,
  M.~Sadkowska-Todys, M.~Rosinska, and W.~J. Edmunds.
\newblock Social contacts and mixing patterns relevant to the spread of
  infectious diseases.
\newblock {\em PLoS Medicine}, 5(3):381--391, Jan 2008.

\bibitem{Newman2002}
M.~Newman.
\newblock {Spread of epidemic disease on networks}.
\newblock {\em Phys. Rev. E}, 66(1), July 2002.

\bibitem{VanKampen1992}
\nooptsort{Kampen} N. G.~van Kampen.
\newblock {\em {Stochastic Processes in Physics and Chemistry}}.
\newblock Elsevier, 1992.

\bibitem{Pastor-Satorras:2001}
R.~Pastor-Satorras and A.~Vespignani.
\newblock Epidemic dynamics and endemic states in complex networks.
\newblock {\em Physical Review E}, 63(066117), 2001.

\bibitem{ARand1999}
D.~A. Rand.
\newblock {Correlation Equations and Pair Approximations for Spatial
  Ecologies}.
\newblock {\em Advanced Ecological Theory}, 12(3-4):100--142, 1999.

\bibitem{Riley:2003}
S.~Riley, C.~Fraser, C.~A. Donnelly, A.~C. Ghani, L.~J. Abu-Raddad, A.~J.
  Hedley, G.~M. Leung, L.-M. Ho, T.-H. Lam, T.~Q. Thach, P.~Chau, K.-P. Chan,
  S.-V. Lo, P.-Y. Leung, T.~Tsang, W.~Ho, K.-H. Lee, E.~M.~C. Lau, N.~M.
  Ferguson, and R.~M. Anderson.
\newblock Transmission dynamics of the etiological agent of {{SARS in Hong
  Kong}}: impact of public health interventions.
\newblock {\em Science}, 300(5627):1961--6, Jun 2003.

\bibitem{Rogers}
T.~Rogers.
\newblock Maximum-entropy moment-closure for stochastic systems on networks.
\newblock {\em Journal of Statistical Mechanics: Theory and Experiment}, 2011.

\bibitem{VRoss2006}
J.~V. Ross.
\newblock {A stochastic metapopulation model accounting for habitat dynamics.}
\newblock {\em J. M. Bio}, 52(6):788--806, 2006.

\bibitem{Schneeberger:2004}
A.~Schneeberger, C.~H. Mercer, S.~A.~J. Gregson, N.~M. Ferguson, C.~A.
  Nyamukapa, R.~M. Anderson, A.~M. Johnson, and G.~P. Garnett.
\newblock Scale-free networks and sexually transmitted diseases: a description
  of observed patterns of sexual contacts in {{Britain and Zimbabwe}}.
\newblock {\em Sexually Transmitted Diseases}, 31(6):380--7, Jun 2004.

\bibitem{Tildesley:2006}
M.~J. Tildesley, N.~J. Savill, D.~J. Shaw, R.~Deardon, S.~P. Brooks, M.~E.~J.
  Woolhouse, B.~T. Grenfell, and M.~J. Keeling.
\newblock Optimal reactive vaccination strategies for a foot-and-mouth outbreak
  in the {UK}.
\newblock {\em Nature}, 440(7080):83--6, Mar 2006.

\bibitem{Volz2008}
E.~Volz.
\newblock {SIR dynamics in random networks with heterogeneous connectivity.}
\newblock {\em J. M. Bio}, 56(3):293--310, 2008.

\bibitem{Volz_Myers}
E.~Volz and L.~A. Myers.
\newblock Susceptible-infected-recovered epidemics in dynamic contact networks.
\newblock {\em Proc. R. Soc. B}, 274(1628):2925--2933, Dec 2007.

\end{thebibliography}

\providecommand{\nooptsort}[1]{}

\newpage

\newpage

\begin{figure}[H]
\centering
\subfloat[]{\includegraphics[width = 0.95\textwidth]{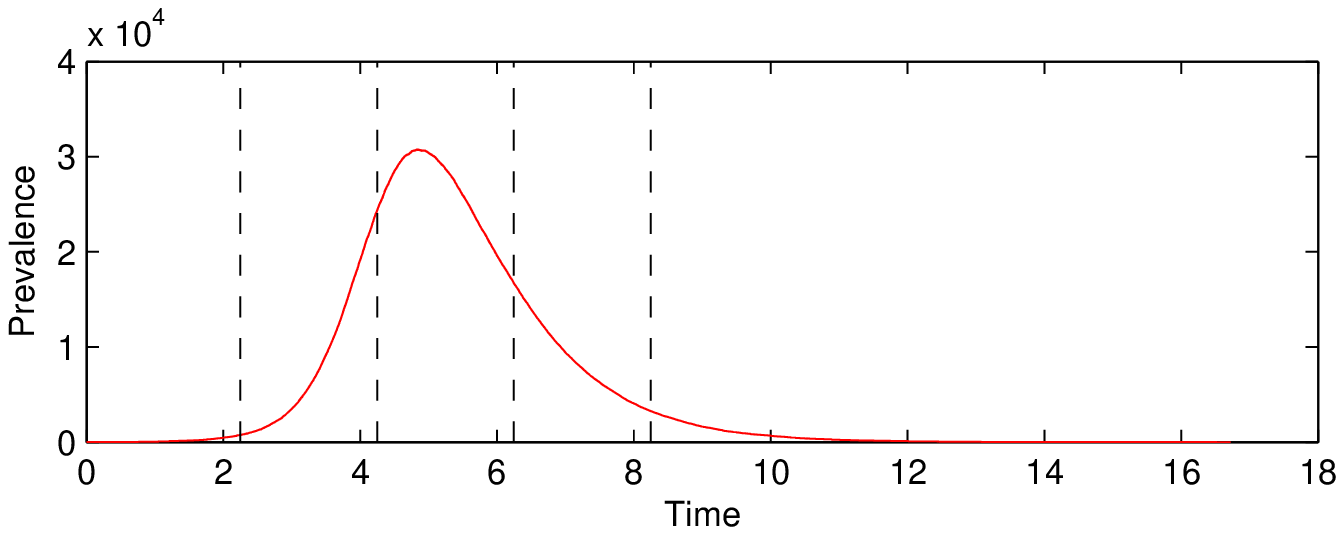}}\\
\subfloat[]{\includegraphics[width = 0.95\textwidth]{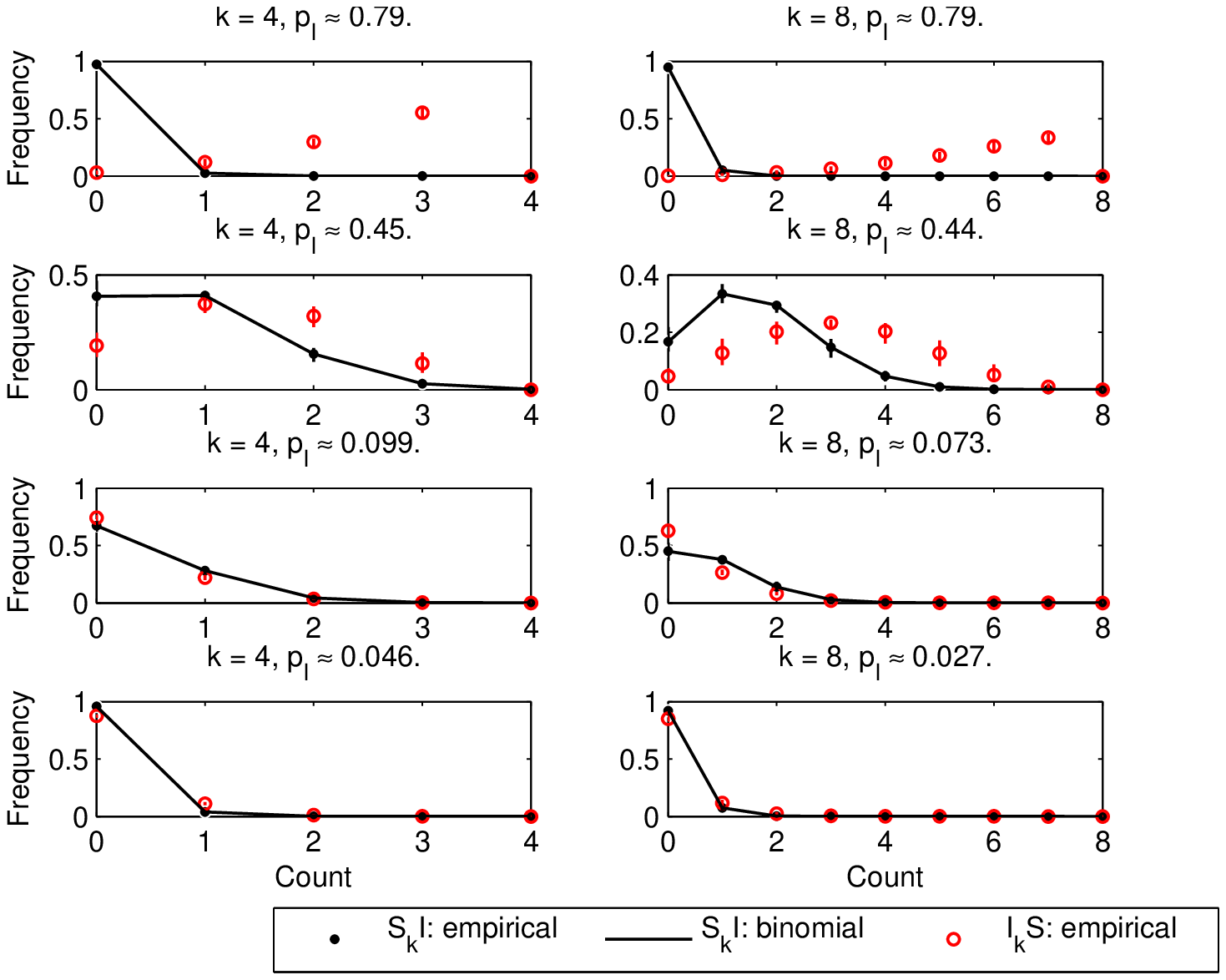}}
\caption{\label{fig:bintest}Test of the binomial assumption. 100 Monte Carlo
realisations were run for parameters $N=10^5$, $P(4) = 2/3$, $P(8) = 1/3$,
$\tau = 0.8$, $\gamma = 1$. (a) shows a typical epidemic curve and four time
points sampled.  (b) shows the empirical histograms for $I$ around an $S_k$
node and $S$ around an $I_k$ node at each time point, and the equivalent
binomial for susceptible central nodes, together with 95\% prediction intervals
across simulations (where each simulation is time-shifted to agree on the first
time at which $[I]=100$) to give an indication of finite-size effects. This
shows the accuracy of the asymptotic results even at finite size.}
\label{fig:infected_assumption} \end{figure}

\newpage

\newpage

\appendix

\include{appendix_name}

\section{Full form of Var$(I)$}  

Here we reproduce the entire expression for the variance of the prevalence
of the infection during the early growth phase. The variance of $[I]$ at time $t$ is denoted by $\sigma_{[I](t)}^2$.
\begin{equation}
\begin{split}
 \sigma_{[I](t)}^2 = & \frac{\tilde{I}}{g'^2 \gamma (\gamma + r)(2 \gamma +r)} \bigg(-\gamma \tau  e^{-2 \gamma  t} (e^{t (\gamma +r)}-1) \Big(\frac{g' \tau  (e^{t (\gamma +r)}-1) (g'' (\frac{\gamma +r}{\gamma  +r+\tau }+\tau )+\frac{g''' (\gamma +r)}{\gamma +r+2 \tau }-g' \tau )}{\gamma +r} \\
& +(g''-g') \big(\gamma  g'+\tau   (g''-g')\big)\Big)+(\gamma + r)2 \tau  e^{r t} \Big(-\frac{g' \tau  (1-e^{r t}) (g'' (\frac{\gamma +r}{\gamma  +r+\tau }+\tau )+\frac{g''' (\gamma +r)}{\gamma +r+2 \tau }-g' \tau )}{r} \\
&-\frac{g' \tau  e^{-\gamma  t} (e^{t (\gamma +r)}-1)   (g'' (\frac{\gamma +r}{\gamma +r+\tau }+\tau )+\frac{g''' (\gamma +r)}{\gamma +r+2 \tau }-g' \tau )}{\gamma +r}+e^{-\gamma  t}   (g'-g'') (\gamma  g'+\tau  (g''-g')) \\
&+(g''-g') \big(\gamma  g'+\tau  (g''-g')\big)\Big)-\gamma e^{-2 \gamma  t} \big(\gamma  g'+\tau  (g''-g')\big) \big(\tau  (g''-g') e^{t (\gamma +r)}+\gamma  g'+g' r+g'   \tau -g'' \tau \big)\\
&+g'^2\gamma (\gamma +r)e^{r t} \big(\gamma +\tau  (\frac{g''}{g'}-1)\big)\bigg)
\text{ ,} \label{eqn:Variance_I}
\end{split}  
 \end{equation}
where $g^{(n)} =g^{(n)}(1)$, $\gamma$ and $\tau$ are the recovery and transmission rates respectively,
 $r$ is the exponential rate of the mean early growth given in \eqref{eqn:t_inf_var}, 
 and as in \eqref{eqn:t_inf_var}, $\tilde{I}$ is a constant related to the
prevalence of infection as the early asymptotic behaviour commences.

\section{Full form of $\mathbf{\hat{G}}$}
We reproduce the matrix $\mathbf{\hat{G}}$ here. Note that  $\gamma$, $\tau$ and $g^{(n)}$
are defined in the same way as in Appendix A and $d_1$ is the proportion of nodes in the network which have degree 1.
\vspace{1mm} 
        
 $\mathbf{\hat{G}}_{\theta,\theta} = \tau  (g''-g')/ N^2 d_1 g'^2$,

 \vspace{1mm}

$\mathbf{\hat{G}}_{\theta,[I]} =    \mathbf{\hat{G}}_{[I],[\theta]} =  \tau  (g'-g'')/N g'^2$,
 
 \vspace{1mm}

$\mathbf{\hat{G}}_{\theta,[SS]} = \mathbf{\hat{G}}_{[SS],[\theta]}= 0$ ,   
 
 \vspace{1mm}

$\mathbf{\hat{G}}_{\theta,[SI]} = \mathbf{\hat{G}}_{[SI],[\theta]} =  \tau  (g''-g')/Ng'^2$,
  
\vspace{1mm}

$\mathbf{\hat{G}}_{[I],[I]} = (\gamma -\tau +\frac{\tau  g''}{g'})$,

  \vspace{1mm}
  
 $\mathbf{\hat{G}}_{[I],[SS]} =\mathbf{\hat{G}}_{[SS],[I]}  = 2  \tau  (g'-g'') g''/g'^2 $,
 
  \vspace{1mm}

   $\mathbf{\hat{G}}_{[I],[SI]} =   \mathbf{\hat{G}}_{[SI],[I]} =       (g'-g'') ((\gamma -\tau ) g'+\tau  g'')/g'^2$,
 
     \vspace{1mm}

$\mathbf{\hat{G}}_{[SS],[SS]} =  -4  \tau  (g'-g'') (g''+g''')/g'^2 $,
 
 \vspace{1mm}

$\mathbf{\hat{G}}_{[SS],[SI]} = 2  \tau    (g'-g'') g'''/g'^2$,
 
  \vspace{1mm}

  $\mathbf{\hat{G}}_{[SI],[SI]}  = \big (-g' \tau +g'' (\tau +(\gamma +r)/(\gamma +r+\tau)  )+g''' (\gamma +r)/(\gamma +r+2 \tau )\big)/g' $.
  
  \section{Branching process approximation}

\label{sec:bp}
  
An alternative approach to modelling the early epidemic is to imbed a branching
process in the full dynamics, which then allows us to generate results relating
to the variance we would observe in the $k$-th generation of infecteds. We let
the degree distribution be $D$, so that the probability that a neighbour of
yours has degree $k$ is given by $k \mathbb{P}(D=k)/\mathbb{E}[D]$, where
$\mathbb{P}(D=k) = d_k$ and $\mathbb{E}(D) = \bar{n}$. Then we consider a
Reed-Frost epidemic on the configuration model of the network given by our
degree distribution. This is a discrete time model, meaning that we are
concerned with the number of invectives in successive ``generations'' of the
disease. We note that if you are an initial infective, then you are in
generation 0 of the epidemic, if you are infected by an initial infective, then
you are in generation 1 and so on.  For a Reed-Frost epidemic we denote the
probability that infection will be spread across an edge from an infective to a
susceptible by $p$. For example, if in generation $n$ you consider an infected
node, then there is an independent probability of $p$, that each of its
neighbours will be an infective of in generation $n+1$.
  
	Suppose that we have $I_n$ infected individuals in generation $n$. Then in
generation $n+1$, there will be $\sum_{i=1}^{I_n} B_i$ infectives, where $B_i$
is the number of `children' of each infective $i$.  Clearly the value $B_i$ is
dependent on the degree of infective $i$, which is say $k$. It is also clear
that this will be at most $k-1$, as to be infective in the first place, one of
the links to our node must have passed the infection down it, meaning that it
has a non-susceptible neighbour at the other end of this link. In this
construction, during the early growth period there will have been no
significant susceptible depletion and therefore we will consider all the
remaining $k-1$ nodes to still be susceptible.
  
  The $B_i$'s are independent and identically distributed. We want to calculate the probability generating function for the $B$'s. To do this
  we calculate,
  \begin{equation}
  \mathbb{E}(s^B) = \sum_{b=0}^{\infty} s^b \mathbb{P} (B=b) \text{. }
  \end{equation}
 To calculate $\mathbb{P}(B=b)$, we use the law of total probability as follows:
 \begin{equation}
  \begin{aligned}
  \mathbb{P}(B=b)& = \sum_{k=0}^{\infty} \mathbb{P}(B=b | D= k) \mathbb{P}(D = k) \text{,} \\
  & = \sum_{k=0}^{\infty} \binom{k-1}{b} p^b (1-p)^{k-1-b} k \frac{d_k}{\mathbb{E}(D)} \text{,}
  \end{aligned}
  \end{equation}
  where we have the $k-1$ in the binomial coefficient and in the powers, to account for the fact that the $k$-th link was responsible for
  the infection in the first place.
  
  So we have,
  \begin{equation}
  \begin{aligned}
  \mathbb{E}(s^B) & = \sum_{b=0}^{\infty} s^b \sum_{k=0}^{\infty} \binom{k-1}{b} p^b (1-p)^{k-1-b} k \frac{d_k}{\mathbb{E}(D)} \text{, } \\
  & = \frac{1}{\mathbb{E}(D)} \sum_{b=0}^{\infty} \sum_{k=0}^{\infty} \binom{k-1}{b} (s p)^b (1-p)^{k-1-b} k d_k \text{.} 
  \end{aligned}
  \end{equation}
  Now we swap the order of summation to get,
  \begin{equation}
  \begin{aligned}
  \mathbb{E}(s^B) & = \frac{1}{\mathbb{E}(D)} \sum_{k=0}^{\infty} \Bigg ( \sum_{b=0}^{k-1} \binom{k-1}{b} (s p)^b (1-p)^{k-1-b} \Bigg ) k d_k \text{,} \\
  & = \frac{1}{\mathbb{E}(D)} \sum_{k=0}^{\infty} k d_k (1-p +sp)^{k-1} \text{ ,} \\
  & =  \frac{1}{\mathbb{E}(D)} g' (1-p+sp) \text{ . }
  \end{aligned}
  \end{equation}
  The expectation of $B$, is then given by taking the derivative of this expression with respect to $s$ and then setting $s$ to 1. When we do this we get,
  \begin{equation}
  \mathbb{E}(B) = \frac{1}{\mathbb{E}(D)}g''(1) = \frac{\mathbb{E}(D(D-1))p}{\mathbb{E}(D)} \text{ . } \label{eqn:expectation_B}
  \end{equation}
  Similarly we get
  \begin{equation}\text{var}(B) = \frac{\mathbb{E}(D(D-1)(D-2))p^2}{\mathbb{E}(D)}  + \frac{\mathbb{E}(D(D-1)) p}{\mathbb{E}(D)} - \bigg( \frac{\mathbb{E}(D(D-1)) p}{\mathbb{E}(D)} \bigg)^2 \text{ .} \label{eqn:var_B}
  \end{equation}
  The expectation for the number of invectives in generation $k$ follows
  \begin{equation}
  \mathbb{E}(I_k) = \mathbb{E}\sum_{i=1}^{I_k-1} B_i = \mathbb{E}(B)\mathbb{E}(I_{k-1}) = \mathbb{E}(B)^k I_0 \text{ . }
  \end{equation}
  The law of total variance then tells us that the variance of $I_k$ is given by,
  \begin{equation}
  \begin{aligned}
  \text{var}(I_k) &= \mathbb{E}\bigg(\text{var}\bigg( \sum_{i=1}^{I_{k-1}} B_i | I_{k-1} \bigg)\bigg) + \text{var}\bigg(\mathbb{E} \sum_{i=1}^{I_{k-1}} B_i | I_{k-1}\bigg) \text{ ,} \\
  & = \mathbb{E} \bigg( \sum_{i=1}^{I_{k-1}} \text{var} (B_i) \bigg) + \text{var}(\mathbb{E}(B_1) I_{k-1}) \text{ ,}\\
  & = \mathbb{E}(I_{k-1})\text{var}(B) + \text{var}(I_{k-1}) \mathbb{E}(B)^2 \text{ ,} \\
  & = I_0 \mathbb{E}(B)^{k-1} \text{var}(B) + \text{var}(I_{k-1}) \mathbb{E}(B)^2 \text{ .}
  \end{aligned}
  \end{equation}
  Using induction this gives us
  \begin{equation}
  \begin{aligned}
  \text{var} (I_k)& = I_0 \mathbb{E}(B)^{k-1}(1+\mathbb{E}(B) + \ldots +\mathbb{E}(B)^{k-1}) \text{var}(B) \text{ , } \\
  & = I_0 \mathbb{E}(B)^{k-1}\frac{\mathbb{E}(B)^{k} -1}{\mathbb{E}(B)-1} \text{var}(B) \text{ .}
  \end{aligned}
  \end{equation}
  When we substitute in the the expression for the mean and variance of $B$ from  \eqref{eqn:expectation_B}  and \eqref{eqn:var_B} respectively, we get,
    \begin{equation}
    \begin{split}
  \text{var} (I_k) = & I_0 \bigg( \frac{\mathbb{E}(D(D-1))p}{\mathbb{E}(D)}\bigg)^{k-1} \bigg(\bigg(\frac{\mathbb{E}(D(D-1))p}{\mathbb{E}(D)}\bigg)^{k} -1\bigg)  \frac{1}{\frac{\mathbb{E}(D(D-1))p}{\mathbb{E}(D)}-1} \\
  & \bigg(\frac{\mathbb{E}(D(D-1)(D-2))p^2}{\mathbb{E}(D)}  + \frac{\mathbb{E}(D(D-1)) p}{\mathbb{E}(D)} - \bigg( \frac{\mathbb{E}(D(D-1)) p}{\mathbb{E}(D)} \bigg)^2 \bigg) \\
=& I_0 \bigg( \frac{g''p}{g'}\bigg)^{k-1} \bigg(\bigg(\frac{g''p}{g'}\bigg)^{k} -1\bigg)  \frac{1}{\frac{g''p}{g'}-1} 
  \bigg(\frac{g'''p^2}{g'}  + \frac{g'' p}{g'} - \bigg( \frac{g'' p}%
{g'} \bigg)^2 \bigg)\text{ .}
  \end{split}
  \end{equation}
Now, letting $k$ become large, and setting $\mathbb{E}(B)^k = e^{rt}$, $p =
\tau/(\tau + \gamma)$, to correspond as closely as possible to continuous-time
results, we obtain
\begin{equation}
\text{var}(I_k) \rightarrow I_0 e^{2rt}
\frac{g' g''' \tau^2  + g' g'' \tau(\tau+\gamma) - (g'')^2 \tau^2}%
{g'(\tau+\gamma)(\tau g''-(\tau+\gamma)g')} \text{ ,}
  \end{equation}
	which is not in agreement with~\eqref{eqn:t_inf_var}. This is as we would
expect, because the reasoning in \S{}\ref{sec:diffmod} above depends critically
on the Markovian nature of the dynamics. In contrast, the branching process can
account for non-Markovian dynamics, but does not allow for the fact that
high-degree infective nodes create new infections more quickly than low-degree
infective nodes. The two approaches are therefore best seen as complementary.

       \end{document}